\documentclass[reprint,superscriptaddress,amsmath,amssymb,aps,pra]{revtex4-1}
\usepackage{amssymb,amsmath}
\usepackage{mathrsfs}
\usepackage{braket}
\usepackage{color}
\usepackage{graphicx}
\usepackage{dcolumn}
\usepackage{bm}

\usepackage[hypertexnames=false,colorlinks=true,linkcolor=red,citecolor=blue,urlcolor=blue]{hyperref} 
\usepackage{natbib} 
\usepackage{float}

\begin{document}

\preprint{APS/123-QED}

\title{Magnon squeezing near a quantum critical point in a cavity-magnon-qubit system}


\author{Gang Liu}
\affiliation{Zhejiang Key Laboratory of Micro-Nano Quantum Chips and Quantum Control, and School of Physics,
	       Zhejiang University, Hangzhou 310027, China}

\author{Gen Li}
\affiliation{Zhejiang Key Laboratory of Micro-Nano Quantum Chips and Quantum Control, and School of Physics,
	       Zhejiang University, Hangzhou 310027, China}

\author{Rong-Can Yang}
\affiliation{Zhejiang Key Laboratory of Micro-Nano Quantum Chips and Quantum Control, and School of Physics,
	       Zhejiang University, Hangzhou 310027, China}

\author{Wei Xiong}%
\email{xiongweiphys@wzu.edu.cn}
\affiliation{Department of Physics, Wenzhou University, Zhejiang, 325035, China}
\affiliation{International Quantum Academy, Shenzhen,518048,China}

\author{Jie Li}
\email{jieli007@zju.edu.cn}
\affiliation{Zhejiang Key Laboratory of Micro-Nano Quantum Chips and Quantum Control, and School of Physics,
	       Zhejiang University, Hangzhou 310027, China}

\date{\today}

\begin{abstract}
	Preparing magnon nonclassical states is a central topic in the study of quantum magnonics. Here we propose to generate magnon squeezed states in a hybrid cavity-magnon-qubit system by engineering an effective Rabi-type magnon-qubit interaction. This is achieved by adiabatically eliminating the cavity mode and driving the qubit with two microwave fields, of which the driving frequencies and amplitudes are properly selected. 
	 {In the normal phase of the effective Rabi model associated with the ground-state superradiant phase transition, the driven magnon–qubit dynamics generate a parametric-amplification-like two-magnon interaction that produces magnon squeezing. The resulting squeezing is strongly enhanced when the system is tuned close to the critical point.}	We further analyze the effects of the dissipation, dephasing, and thermal noise on the magnon squeezing.  Our results indicate that a moderate degree of squeezing can be produced using currently available parameters in the experiments.
\end{abstract}

\maketitle


\section{\label{sec:level1}INTRODUCTION}

Hybrid quantum systems based on magnonics have attracted considerable attention over the past decade due to their potential for enabling various applications in quantum information science, quantum sensing, quantum technology, and macroscopic quantum physics~\cite{lachance2019hybrid,yuan2022quantum,Rameshti-2022,zuo2024cavity}. Magnons are quantized excitations of a spin wave in magnetically ordered systems, such as yttrium iron garnet (YIG). The magnon mode in a YIG sample exhibits many excellent properties, including high spin density, low dissipation rate, and great tunability. These features render YIG-based magnonic systems particularly attractive for realizing coherent coupling with diverse quantum systems and preparing magnonic quantum states~\cite{lachance2019hybrid,zuo2024cavity}. Indeed, magnon modes have been demonstrated to coherently interact with microwave photons via the magnetic dipole coupling~\cite{Huebl2013prl,Tabuchi2014prl,Zhang2014prl}, optical photons via the magneto-optical effect~\cite{hisatomi2016bidirectional,osada2016cavity,zhang2016optomagnonic,Haigh2016prl}, and mechanical vibrations via the magnetostrictive interaction~\cite{zhang2016cavity,li2018magnon,Potts2021prx,shen2022prl,shen2025nc}. Furthermore, magnons can also couple to superconducting qubits through a common microwave cavity~\cite{tabuchi2015coherent,lachance2017resolving,lachance2020entanglement,wolski2020dissipation,xu2023quantum}. 
Recent experiments indicate that the hybrid cavity-magnon-qubit system can be an ideal platform to realize the generation and quantum control of magnonic quantum states. Important experimental breakthroughs include the realization of magnon-qubit strong coupling~\cite{tabuchi2015coherent}, single-shot detection of a single magnon~\cite{lachance2020entanglement}, and magnonic superposition states~\cite{xu2023quantum}. These achievements have stimulated a growing number of theoretical proposals for the generation and manipulation of magnonic nonclassical states, such as magnon blockade~\cite{liu2019prbblockade,xie2020prablockade,wu2021phase,wang2022dissipation,Jin2023prablockade}, entangled states~\cite{kong2021,qi2022generation,ren2022long,wang2022,lu2025}, squeezed states~\cite{2023guoprasqueezed,xia2025magnon}, and cat states~\cite{Kounalakis2022prl,he2023pramc,hou2024robust,he2024pracats,liu2025magnoncat}.

In recent years, a number of proposals have been offered to prepare magnon squeezed states, e.g., by transferring squeezing from a microwave field~\cite{li2019squeezed} or a phonon mode~\cite{fan2024magnon} to the magnon mode, two-tone driving of the magnon mode~\cite{zhang2021squeezed,qian2024squeezing}, exploiting the anisotropy of the ferromagnet~\cite{Kamra2016prl_squeezed} and the mechanism of the ponderomotivelike squeezing~\cite{li2022nsr_squeezing}, 
and using an easy-axis ferromagnet~\cite{Lee2023squeezing}. 
In cavity-magnon-qubit systems, magnon squeezing can be achieved by two-tone driving of a superconducting qubit to induce the magnonic parametric amplification~\cite{2023guoprasqueezed}, and engineering a nonlinear magnon-qubit coupling~\cite{xia2025magnon}.
 {We note that the pulse protocol~\cite{sharma2022protocol} has a potential to prepare a magnon squeezed state, but due to the very short magnon lifetime and limited magnon-qubit coupling strength in the practical situation~\cite{lachance2020entanglement,xu2023quantum}, the protocol is limited to generate a superposition state of Fock states $|N \rangle$ up to $N=3$, which is incompetent to construct a squeezed vacuum state  that corresponds to an infinite dimensional Hilbert space. }

Here, we propose a different mechanism for achieving magnon squeezing by engineering a Rabi-type magnon-qubit interaction in the cavity-magnon-qubit system.  Specifically, when the microwave cavity is far detuned from both the magnon and qubit system, the cavity mode can be adiabatically eliminated, yielding an effective magnon-qubit coupling mediated by virtual photons. By driving the superconducting qubit with two microwave fields and appropriately adjusting the drive frequencies and amplitudes, we show that the magnon-qubit system can realize an effective quantum Rabi model~\cite{prx2012simulation,braumuller2017analog}.
 {Notably, in the normal phase the effective magnon–qubit dynamics already contain a parametric-amplification-like two-magnon interaction for the magnon mode. This interaction generates magnon squeezing and becomes increasingly strong as the system is tuned towards the critical point associated with the ground-state superradiant phase transition.}
Using experimentally accessible parameters, our scheme can achieve a magnon squeezing of about  {3.7} dB. The generated squeezing is robust against both the system dissipations and thermal noise. 
Compared to the existing qubit-assisted schemes~\cite{2023guoprasqueezed,xia2025magnon}, our approach features a distinct underlying mechanism.   {Specifically, Ref.~\cite{2023guoprasqueezed} achieves magnon squeezing through direct parametric amplification within a driven effective Jaynes-Cummings (JC) model. By contrast, our scheme critically relies on the phase transition of an effective quantum Rabi model.   The squeezing mechanism of the Ref.~\cite{xia2025magnon} is based on an engineered nonlinear coupling between the superconducting qubit and the microwave resonator, which is however only present for a microwave transmission line resonator but absent for a 3D microwave cavity. }

 
The paper is organized as follows: In Sec.~\ref{sec:level2}, we describe our model of the cavity-magnon-qubit hybrid system and derive the effective Hamiltonian of the magnon-qubit system with the presence of two drive fields, which can yield magnon squeezing. In Sec.~\ref{sec:level3}, we derive the master equation to include the dissipation and noise effects, present the results of the magnon squeezing, and analyze the effects of the magnon and qubit dissipations and bath temperature on the squeezing. Finally, we summarize our findings in Sec.~\ref{sec:level4}.

\begin{figure}[t]
	\centering
	\includegraphics[width=0.96\linewidth]{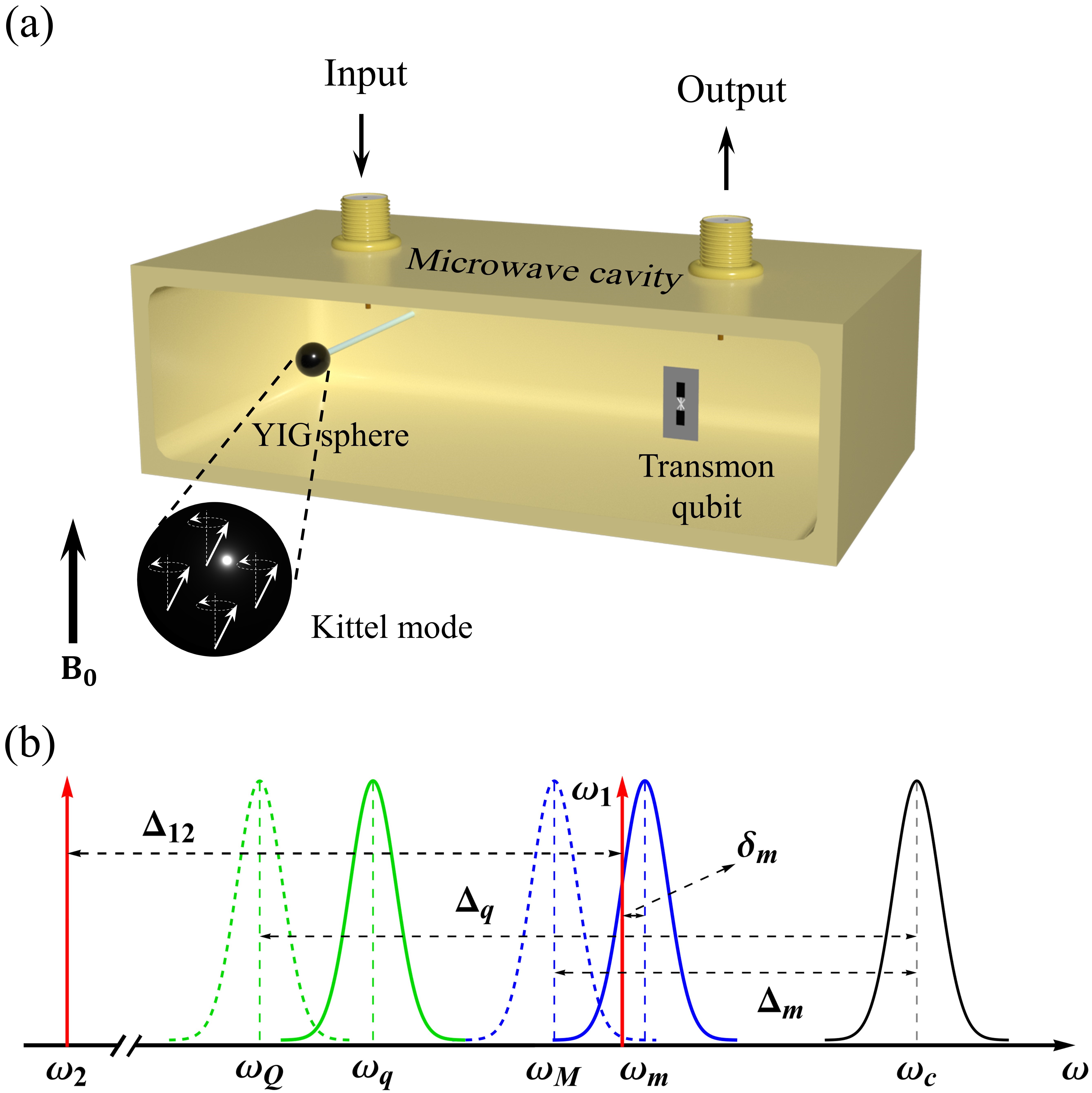}
	\caption{(a) Schematic of the cavity-magnon-qubit system. A microwave cavity mode simultaneously couples to a magnon mode of a YIG sphere via the magnetic dipole interaction and to a superconducting qubit via the electric dipole interaction. The YIG sphere is placed inside a bias magnetic field $B_0$, and the qubit is further driven by two microwave fields. An effective magnon-qubit interaction is mediated by the off-resonant cavity mode. (b) Mode and drive frequencies of the system. The cavity mode with frequency $\omega_c$ is detuned from both the magnon mode with frequency  $\omega_M$ and the qubit with frequency $\omega_Q$. The elimination of the cavity mode leads to the effective frequencies of the magnon mode and the qubit, denoted as $\omega_m$ and $\omega_q$, respectively. See text for the details of two drive fields (with frequencies $\omega_1$ and $\omega_2$) applied to the qubit. }
	\label{fig:fig1}
\end{figure}

\section{\label{sec:level2}The system and the effective magnon-qubit Hamiltonian}

The system under study consists of a YIG sphere supporting a magnon mode (e.g., the Kittel mode), a transmon superconducting qubit, and a three-dimensional microwave cavity, as illustrated in Fig.~\ref{fig:fig1}(a). The YIG sphere is positioned inside a uniform static magnetic field $B_0$ and the superconducting qubit is driven by two microwave fields. The magnon mode and the superconducting qubit are simultaneously and strongly coupled to the microwave cavity via the magnetic dipole and the electric dipole interactions, respectively, with the coupling strengths denoted as $g_{\rm cm}$ and $g_{\rm cq}$.  Thus, the Hamiltonian of the cavity-magnon-qubit system is given by
\begin{align}\label{eq1}
	H/\hbar =&\ \omega_c c^\dagger c + \frac{1}{2}\omega_Q \sigma_z + \omega_M m^\dagger m \notag\\
	& + g_{\rm cq} (c \sigma_+ + c^\dagger \sigma_-)+ g_{\rm cm} (c m^\dagger + c^\dagger m),
\end{align}
where the annihilation (creation) operators $c$~($c^\dagger$) and $m$~($m^\dagger$) are of the cavity and magnon modes, respectively, whose resonance frequencies are $\omega_c$ and $\omega_M$.  $\omega_Q$ is the transition frequency of the superconducting qubit with the associated lowering (raising) operator $\sigma_-=|g\rangle\langle e|$~($\sigma_+=|e\rangle\langle g|$). Here, $|g\rangle$~($|e\rangle$) denotes the ground (excited) state of the qubit, and $\sigma_z = |e\rangle\langle e| - |g\rangle\langle g|$ is the Pauli-$z$ operator.

We consider the situation where both the magnon mode and the superconducting qubit are far detuned from the cavity mode, i.e.,
$\Delta_q = \omega_c - \omega_Q \gg g_{\rm cq}$, and $\Delta_m = \omega_c - \omega_M \gg g_{\rm cm}$. As a result, the cavity mode is virtually excited and can be adiabatically eliminated. This gives rise to an effective JC-type coupling between the superconducting qubit and the magnon mode, i.e.,
\begin{align}
	H_{\mathrm{JC}}/\hbar = \omega_m m^\dagger m + \frac{1}{2}\omega_q \sigma_z + G(\sigma_+ m + \sigma_- m^\dagger),\label{eq2}
\end{align}
where $\omega_m = \omega_M  + g_{\rm cm}^2/\Delta_m$ is the effective frequency of the magnon mode, $\omega_q = \omega_Q + g_{\rm cq}^2/\Delta_q$ is the effective frequency of the superconducting qubit, as depicted in Fig. \ref{fig:fig1}(b), and $G = {g_{\rm cq} g_{\rm cm}} (\Delta_m^{-1} + \Delta_q^{-1})/2$ is the effective coupling strength between the qubit and the magnon mode.  

We then apply two microwave fields to drive the superconducting qubit, and the drive frequencies are $\omega_1$ and $\omega_2$. In the frame rotating at the frequency $\omega_1$, by making the transformation $\sigma_+\rightarrow \sigma_+ e^{i\omega_1 t}$, $\sigma_-\rightarrow \sigma_- e^{-i\omega_1 t}$, $m^\dagger\rightarrow m^\dagger e^{i\omega_1 t}$, and $m\rightarrow m e^{-i\omega_1 t}$, the Hamiltonian \eqref{eq2} becomes 
\begin{align}\label{eq:jcprime}
	H_{\mathrm{JC}}/\hbar =&\ \delta_m m^\dagger m + \frac{1}{2}\delta_q \sigma_z 
	                         + G(\sigma_+ m + \sigma_- m^\dagger)\nonumber\\
						   & + \mathcal{E}_1 (\sigma_+  + \sigma_- )
						     + \mathcal{E}_2 (\sigma_+ e^{i\Delta_{12}t}+\sigma_- e^{-i\Delta_{12} t}),
\end{align}
where $\delta_{m(q)} = \omega_{m(q)} - \omega_1$ denotes the detuning of the magnon mode (the qubit) with respect to the drive field of $\omega_1$, and $\Delta_{12} = \omega_1 - \omega_2$ is the detuning between the two drive fields. Here, $\mathcal{E}_1$ and $\mathcal{E}_2$ are the Rabi frequencies associated with the two drive fields. For simplicity, we assume $\mathcal{E}_1$ and $\mathcal{E}_2$ to be real.  The effect of the relative phase between the two drive fields is analyzed later in Appendix~\ref{appendixA}.

Under the interaction with the drive field of $\omega_1$, the superconducting qubit is dressed with the eigenstates $\ket{\pm} = (\ket{g}\pm \ket{e})/\sqrt{2}$. Under the dressed states, we have the transformations $\sigma_x\rightarrow \tau_z$ and $\sigma_z\rightarrow-\tau_x$, where $\tau_z=|+\rangle\langle+|-|-\rangle\langle-|$ and $\tau_x=|+\rangle\langle-|+|-\rangle\langle+|$. Hence, the Hamiltonian~\eqref{eq:jcprime} can be rewritten as
\begin{align}\label{eq4}
	H_{\mathrm{JC}}/\hbar =& \ \delta_m m^\dagger m  - \frac{\delta_q}{2} \tau_x  \notag \\
	&+\frac{G}{2}\left[(m+m^\dagger)\tau_z+i(m-m^\dagger)\tau_y\right] \notag \\
	&+\mathcal{E}_1\tau_z+\mathcal{E}_2 \big[\tau_z\cos(\Delta_{12}t)-\tau_y\sin(\Delta_{12}t) \big],
\end{align}
where $\tau_y=i(|-\rangle\langle+|-|+\rangle\langle-|)$. Moving into the rotating frame with respect to the Rabi frequency $\mathcal{E}_1$, the system Hamiltonian \eqref{eq4} becomes
\begin{align}\label{eq5}
		H_{\mathrm{JC}}/\hbar =& \ \delta_m m^\dagger m +\mathcal{E}_2\tau_z\cos(\Delta_{12}t)\notag\\
		& - \frac{\delta_q}{2}\big[\tau_x\cos(2\mathcal{E}_1t) - \tau_y\sin(2\mathcal{E}_1t) \big]\notag\\
		&-\mathcal{E}_2 \sin(\Delta_{12}t) \big[\tau_x\sin(2\mathcal{E}_1t) + \tau_y\cos(2\mathcal{E}_1t) \big]\notag\\
		&+\frac{i}{2}G(m-m^\dagger) \big[\tau_x\sin(2\mathcal{E}_1t) + \tau_y\cos(2\mathcal{E}_1t) \big]\notag\\
		&+\frac{1}{2}G(m+m^\dagger)\tau_z.
\end{align}
By setting the detuning of the two drive fields $\Delta_{12}=2\mathcal{E}_1$, and working in the regime $\Delta_{12} \gg \mathcal{E}_2,G$, which validates the rotating-wave approximation (RWA), the time-dependent fast oscillating terms in Eq.~\eqref{eq5} can be neglected, leading to the following Hamiltonian: 
\begin{align}
	H_{\mathrm{Rabi}}/\hbar = \delta_m m^\dagger m - \frac{\mathcal{E}_2}{2}\tau_x + \frac{1}{2}G(m+m^\dagger)\tau_z.	
\end{align}
The above Hamiltonian corresponds to the well-known Rabi model. To cast it into the standard form, we rewrite it in terms of the eigenstates of the operator $\sigma_z$, which then becomes 
\begin{align}\label{eq:rabi}
	H_{\mathrm{Rabi}}/\hbar = \delta_m m^\dagger m + \frac{\mathcal{E}_2}{2}\sigma_z + g(m+m^\dagger)\sigma_x,
\end{align}
with $g=G/2$.  {The validity of the RWA used to derive Eq.~\eqref{eq:rabi} is explicitly analysed in Appendix~\ref{appendixB}.}
 This Hamiltonian can capture the dynamics and rich physics in the ultrastrong ($0.1 < g/\delta_m < 1$) and even deep-strong ($g/\delta_m \gtrsim 1$) coupling regimes~\cite{frisk2019ultrastrong,forn2019ultrastrong}.

In the limit $\zeta=\delta_m/\mathcal{E}_2 \rightarrow 0$, the system described by the Hamiltonian \eqref{eq:rabi} undergoes a quantum phase transition at the critical point $g_c \equiv 2g/\sqrt{\mathcal{E}_2 \delta_m}=1$. Specifically, as the dimensionless coupling strength $g_c$ increases across the critical point $g_c = 1$, the system evolves from the normal phase to the superradiant phase. In the normal phase, the Hamiltonian  \eqref{eq:rabi}  can be analytically diagonalized~\cite{hwang2015quantum,lu2018entanglement,liu2023switchable,liu2025qpt}. The diagonalization can be achieved by applying a unitary transformation $U={\rm exp}[-i(g/\mathcal{E}_2)(m^\dagger+m)\sigma_y ]$ to Eq.~\eqref{eq:rabi}, which yields
\begin{align}\label{eq:hrabi3}
	H_{\mathrm{eff}}/\hbar = U^{\dagger}H_{\text{Rabi}} U 
	= \sum_{k=0}^{\infty}\frac{1}{k!}\left[H_{\text{Rabi}}, V \right]^{(k)},
\end{align}
where the nested commutator is defined in the following way: $\big[H_{\text{Rabi}},V \big]^{(k)} \equiv \left[ \big[H_{\text{Rabi}},V \big]^{(k-1)},V \right]$, with $\big[H_{\text{Rabi}}, V \big]^{(0)} = H_{\text{Rabi}}$. 
Expanding Eq.~\eqref{eq:hrabi3} gives rise to
\begin{align}\label{eq:hrabi32}
	H_{\mathrm{eff}}/\hbar 
	=& \  \delta_m m^\dag m + \frac{\mathcal{E}_2}{2}\sigma_z +  \frac{\delta_m g_c^2}{4}(m^\dag + m)^2 \sigma_z + \mathcal{C} \nonumber\\
	&+ \frac{\delta_m g_c}{2}\sqrt{\frac{\delta_m}{\mathcal{E}_2}}(m^\dag - m)(\sigma_+ - \sigma_-)\nonumber\\
	&+ \frac{\delta_m g_c^3}{6}\sqrt{\frac{\delta_m}{\mathcal{E}_2}}(m^\dag + m)^3 \sigma_x + \mathcal{O}\left(\sqrt{\frac{\delta_m}{\mathcal{E}_2}}\right),
\end{align}
where $\mathcal{C}$ is a constant term and the last term denotes higher than first order terms of $\sqrt{\delta_m/\mathcal{E}_2}$. In the limit $\zeta=\delta_m/\mathcal{E}_2 \rightarrow 0$, the terms in the second and third lines of Eq.~\eqref{eq:hrabi32} can be neglected, resulting in the following approximate form: 
\begin{align}\label{eq:heff}
	H_\text{eff}/\hbar \approx  \delta_m m^\dagger m +  \frac{\mathcal{E}_2}{2}\sigma_z +  \frac{\delta_m g_c^2}{4}(m^\dagger+m)^2\sigma_z.
\end{align}
Projecting the above Hamiltonian onto the ground state of the qubit $\ket{g}$, we obtain the following Hamiltonian for the magnon mode
\begin{align}\label{eq:hqua}
	H_\text{eff}^m /\hbar= \bra{g}H_\text{eff}\ket{g} 
	= \delta_m m^\dagger m - \frac{\delta_m g_c^2}{4}(m^\dagger+m)^2.
\end{align}
The above Hamiltonian describes a parametric amplification-like two-magnon process, which we show later can induce the squeezing of the magnon mode.

\begin{figure*}
	\centering
	\includegraphics[width=0.95\linewidth]{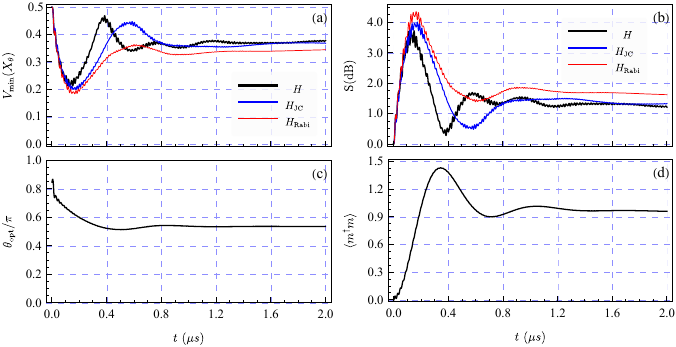}
	\caption{(a) The minimum variance of the quadrature $V_{\rm min}(X_{\theta})$ and (b) the degree of squeezing $S$ (dB) as a function of time, which are obtained at an optimal squeezing angle at each time point.  { In panels (a) and (b), the black curves are the results obtained from the full Hamiltonian in Eq.~\eqref{eq1} including the two-tone driving, the blue thinner curves are obtained from Eq.~\eqref{eq:jcprime} after adiabatic elimination of the cavity in the large-detuning regime, and the red thinnest curves correspond to the Hamiltonian in Eq.~(\ref{eq:rabi}) after the RWA. } 
	(c) The optimal squeezing angle $\theta_{\rm opt}$ and (d) the mean magnon number $\braket{m^\dag m}$ during the evolution, which are obtained using the master equation in Eq.~\eqref{eq:me2}. The parameters adopted are provided in the main text. }
	\label{fig:fig2}
\end{figure*}

\section{\label{sec:level3}Dynamical magnon squeezing}

In the above section, we have analytically shown that our protocol can in principle generate magnon squeezing. In practice, the system suffers from dissipation and dephasing due to the coupling with the external environment, i.e., the thermal bath, which reduce the degree of the squeezing. In what follows, we adopt the master equation approach to numerically study these effects on the magnon squeezing. Note that Eq.~\eqref{eq:rabi} is derived by focusing only on the Hamiltonian transformation. However, a consistent treatment of open quantum systems requires the corresponding master equation to be transformed as well. Accordingly, we derive the master equation for the effective Hamiltonian starting from the Lindblad equation associated with the original full Hamiltonian $H_{\rm JC}$ in Eq.~\eqref{eq:jcprime}, i.e.,
	\begin{align}\label{eq:me}
		\frac{d\rho}{dt} &= -i[H_{\text{JC}}/\hbar, \rho]  + \frac{\kappa}{2}(\bar{n}_{{m}} + 1)\mathcal{L}[m]\rho + \frac{\kappa}{2}\bar{n}_{{m}}\mathcal{L}[m^\dagger]\rho  \nonumber\\
		& +  \frac{\gamma}{2}(\bar{n}_{{q}} + 1)\mathcal{L}[\sigma_-]\rho + \frac{\gamma}{2}\bar{n}_{{q}}\mathcal{L}[\sigma_+]\rho + \frac{\gamma_\phi}{4} \mathcal{L}[{\sigma}_z]\rho ,
	\end{align}
where $\rho$ is the density operator of the system, and  $\mathcal{L}[o]\rho = 2o\rho o^\dagger - o^\dagger o \rho - \rho o^\dagger o$, with the Liouvillian superoperator $\mathcal{L}[o]$, and $o=\{m, {\sigma}_-, {\sigma}_z\}$. Here, $\kappa$~($\gamma$) is the dissipation rate of the magnon mode (the qubit), $\gamma_\phi$ characterizes the pure dephasing rate of the qubit, and $\bar{n}_{m(q)} = [\exp(\hbar \omega_{m(q)}/k_B T) - 1]^{-1}$ is the mean thermal occupation number, with $T$ being the bath temperature and $k_B$ the Boltzmann constant.  After applying the transformations and the RWA introduced in Sec.~\ref{sec:level2}, the master equation in the rotating frame takes the form
	\begin{align}\label{eq:me2}
		\frac{d\tilde{\rho}}{dt} &= -i[H_{\text{Rabi}}/\hbar, \tilde{\rho}]  + \frac{\kappa}{2}(\bar{n}_{{m}} + 1)\mathcal{L}[m]\tilde{\rho} + \frac{\kappa}{2}\bar{n}_{{m}}\mathcal{L}[m^\dagger]\tilde{\rho}\nonumber\\
		& + \frac{\gamma}{4}(2\bar{n}_{{q}} + 1)\mathcal{L}[\sigma_x]\tilde{\rho},
	\end{align}
where $H_{\text{Rabi}}$ corresponds to the one in Eq.~\eqref{eq:rabi} and $\tilde{\rho}$ denotes the density matrix in the rotating frame. We note that the pure dephasing term (last term in Eq.~\eqref{eq:me}) does not appear in Eq.~\eqref{eq:me2}, since the $\sigma_z$-related contributions are discarded in the RWA.

In the simulation, the magnon mode is initially prepared in the thermal state, while the superconducting qubit is initialized in the ground state. To characterize the magnon squeezing, we adopt the variance $V(X_\theta)$ of the general quadrature $X_\theta$ of the magnon mode, which is defined as $X_\theta = \cos(\theta) X_1 + \sin(\theta) X_2$, with the amplitude and phase quadratures of the magnon mode defined as $X_1 = (m + m^\dagger)/\sqrt{2}$ and $X_2 = i(m^\dagger - m)/\sqrt{2}$, respectively. The magnon being squeezed means that the variance  $V(X_\theta)$ is below that of the vacuum fluctuation, i.e., $V(X_\theta) <V_{\rm vac}= 0.5$. In what follows, we only present the results corresponding to the minimum variance $V_{\rm min}(X_{\theta})$ at an optimal squeezing angle $\theta_{\rm opt}$.  The degree of squeezing is quantified in units of dB, defined as $S = -10 \log_{10} [V_\text{min}(X_\theta)/V_{\rm vac}]$.

\begin{figure}[t]
	\centering
	\includegraphics[width=0.95\linewidth]{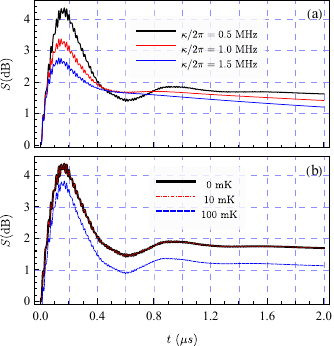}
	\caption{The degree of squeezing $S$ (dB) as a function of time for (a) different dissipation rates of the magnon mode; and (b) different bath temperatures.  We take $T = 10~\text{mK}$ in (a), and $\kappa/2\pi = 0.5~\text{MHz}$ in (b). The other parameters are the same as those in Fig.~\ref{fig:fig2}.  }
	\label{fig:fig3}
\end{figure}

 {
In Figs.~\ref{fig:fig2}(a) and \ref{fig:fig2}(b), we plot the minimum variance $V_{\rm min}(X_{\theta})$ and the degree of squeezing $S$ as a function of time, and compare the results obtained from the full Hamiltonian in Eq.~\eqref{eq1} including the two-tone driving, the effective JC Hamiltonian in Eq.~(\ref{eq:jcprime}) after adiabatic elimination of the cavity, and the effective Rabi-type Hamiltonian in Eq.~(\ref{eq:rabi}) after the RWA. The three models predict qualitatively consistent squeezing dynamics with almost identical time to achieve the maximum squeezing. Their moderate quantitative difference is due to the fundamental inconsistency between the requirements on system parameters for satisfying the condition of the approximation and for achieving a large degree of squeezing.  Specifically, enforcing a strict large-detuning condition improves the accuracy of the adiabatic elimination and the RWA, but yields a reduced effective coupling strength $G$ and thus a weaker squeezing. Relaxing this condition enhances the nonlinear interaction and thus the degree of squeezing but at the cost of the deviation of the dynamics predicted by the three models.}
The results indicate that a magnon squeezing of approximately  {3.7}~dB can be achieved at an optimal time, and the Hamiltonian \eqref{eq:rabi} is a good approximation especially at the time corresponding to the maximum squeezing. 
We use the following experimentally feasible parameters~\cite{tabuchi2015coherent,lachance2017resolving,xu2023quantum}:
$\omega_c/2\pi = {  6.4720} \,\mathrm{GHz}$, 
$\omega_Q/2\pi = { 5.8447} \,\mathrm{GHz}$,
$\omega_M/2\pi = { 5.9205} \,\mathrm{GHz}$, 
$g_{\rm cq}/2\pi = { 85.7} \,\mathrm{MHz}$ and 
$g_{\rm cm}/2\pi = { 72.5} \,\mathrm{MHz}$, which lead to the detunings 
$\Delta_q/2\pi =  { 627.3} \,\mathrm{MHz}$ and 
$\Delta_m/2\pi =  { 551.5 }\,\mathrm{MHz}$,  the effective frequencies  
$\omega_q/2\pi = { 5.85641} \,\mathrm{GHz}$ and 
$\omega_m/2\pi = { 5.93003} \,\mathrm{GHz}$, and the effective coupling strength 
$G/2\pi = { 10.5854} \,\mathrm{MHz}$. 
For the drives, we set 
$\mathcal{E}_1/2\pi = 500~\text{MHz}$, 
$\mathcal{E}_2/2\pi = { 50}~\text{MHz}$, 
$\omega_1/2\pi = { 5.928}~\text{GHz}$, 
$\omega_2/2\pi = { 4.928}~\text{GHz}$, and 
$\delta_m/2\pi = { 2.03083}~\text{MHz}$. 
The other parameters are
 $\kappa/2\pi=0.5$~MHz~\cite{shen2025nc}, $\gamma/2\pi=\gamma_\phi/2\pi=3$~kHz, and $T=10$~mK.  
We note that by using a stronger drive field, yielding a larger $\mathcal{E}_1$, the condition for taking the RWA to get Eq.~(\ref{eq:rabi}) is better satisfied, which will lead the two curves in Figs.~\ref{fig:fig2}(a) and \ref{fig:fig2}(b) to be closer.
%

In Fig.~\ref{fig:fig2}(c), we show the evolution of the optimal squeezing angle, at which the minimum variance $V_{\rm min}(X_{\theta})$ is achieved. The optimal squeezing angle oscillates at an early stage but gradually stabilizes.   It is also interesting to check the mean magnon number during the evolution.  Figure~\ref{fig:fig2}(d) shows that the mean magnon excitation is less than unity throughout the evolution, implying that the magnon-number distribution of the squeezed state is concentrated in a few lowest excited states.	 Both Figs.~\ref{fig:fig2}(c) and \ref{fig:fig2}(d) are plotted using the master equation in Eq.~\eqref{eq:me2}.

\begin{figure}[t]
	\centering
	\includegraphics[width=0.95\linewidth]{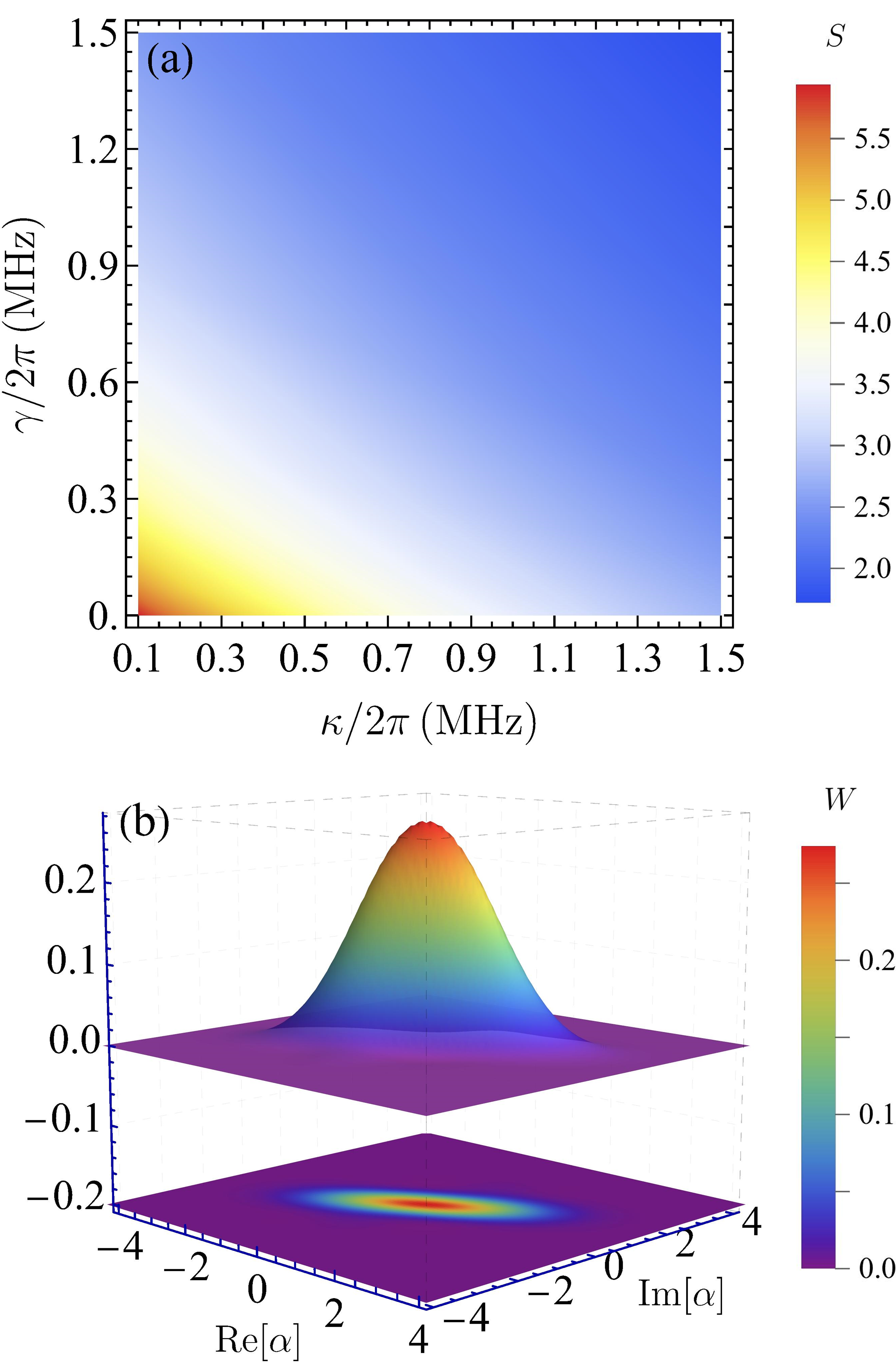}
	\caption{(a) The maximum degree of squeezing $S$ versus the magnon and qubit dissipation rates $\kappa$ and $\gamma$.  Here, the maximum  degree of squeezing  is achieved by optimizing the squeezing angle and evolution time at each value of $\kappa$ and $\gamma$. (b)  The Wigner function of the magnon mode corresponding to the maximum squeezing at an optimal time. 	The other parameters are the same as those in Fig.~\ref{fig:fig2}.		}
	\label{fig:fig4}
\end{figure}

The effect of the magnon dissipation rate on the squeezing is shown in Fig.~\ref{fig:fig3}(a). Clearly, the magnon dissipation can significantly reduce the degree of squeezing, especially at an early stage of the evolution yielding the maximum squeezing. Using one of the lowest damping rates, $\kappa/2\pi \approx 0.5$ MHz, reported in a recent experiment~\cite{shen2025nc} allows us to achieve the squeezing above 4 dB.  
In contrast, at longer evolution times the squeezing becomes less sensitive to the magnon dissipation rate.
Figure~\ref{fig:fig3}(b) shows the influence of the bath temperature on the squeezing. For the temperature below a few tens of mK, the thermal magnon excitation is negligibly small, and the magnon mode is initially in the vacuum state. In this case, the thermal noise is too weak to affect the squeezing, whereas for temperatures above 100 mK, the thermal excitations will have an appreciable effect leading to a reduced degree of squeezing.

In Fig.~\ref{fig:fig4}(a), we show the maximum magnon squeezing versus the magnon and qubit dissipation rates (the effect of the qubit dephasing is negligible under the condition of the RWA shown in Eq.~\eqref{eq:me2}). Clearly, both the dissipation rates have a strong effect on the squeezing. However, due to the fact that a much smaller qubit dissipation rate can be achieved in the experiments \cite{whh2019cat,Ren2022} than the magnon dissipation rate (of a YIG sphere), which is typically around $\kappa/2\pi \sim 1$ MHz because of the intrinsic loss~\cite{Tabuchi2014prl,Zhang2014prl}, the dominant factor that limits our scheme from achieving a much higher degree of squeezing is the considerable magnon dissipation.

 {In the experiment, the squeezing can be verified by performing the Wigner tomography of the magnon mode~\cite{xu2023quantum}. It is essentially to measure the expectation value of the magnon displaced parity operator, i.e., 
\begin{equation}\label{eq11}
	W(\alpha) = \frac{2}{\pi}\text{Tr}\left[e^{i\pi m^\dag m}D^{\dagger}(\alpha){\rho} D(\alpha)\right],
\end{equation}
where $D(\alpha) = \exp(\alpha m^{\dagger} - \alpha^* m)$ denotes the magnon displacement operator with $\alpha$ being a complex variable and $e^{i\pi m^\dag m}=(-1)^{m^\dagger m}$ is the magnon parity operator. }
In Fig.~\ref{fig:fig4}(b), we plot the Wigner function of the magnon mode corresponding to the maximum squeezing at an optimal time, which clearly shows a squeezed magnon mode at an optimal squeezing angle. 

\section{Conclusions}\label{sec:level4}

In summary, we have presented an experimentally feasible scheme for generating magnon squeezed states in a microwave cavity-magnon-qubit system.  {By precisely controlling the frequencies and amplitudes of the two driving fields applied to the qubit, an effective Rabi-type magnon-qubit interaction can be achieved, which yields a parametric-amplification-like two-magnon process that dynamically generates magnon squeezed states in the normal phase. The resulting squeezing can be strongly enhanced when the system is operated close to the ground-state phase-transition point.} Our results indicate that a moderate degree of magnon squeezing about ${ 3.7}$~dB can be achieved using currently available parameters in relevant experiments, which highlight the potential of the qubit-assisted hybrid platforms for engineering magnon  nonclassical states. The magnon squeezed states offer promising opportunities for quantum information processing and quantum sensing with magnons.

\begin{acknowledgments}
We thank Dr. Da Xu for discussions on the experimental feasibility. This research was supported by the Zhejiang Provincial Natural Science Foundation of China (Grant No. LR25A050001), the Natural Science Foundation of Zhejiang Province (Grant No. LY24A040004), the National Natural Science Foundation of China (Grants No. 12474365 and No. 92265202), the National Key Research and Development Program of China (Grants No. 2022YFA1405200 and No. 2024YFA1408900), and the Shenzhen International Quantum Academy (Grant No. SIQA2024KFKT010).
\end{acknowledgments}

\appendix

\begin{figure}[t]
	\centering
	\includegraphics[width=1\linewidth]{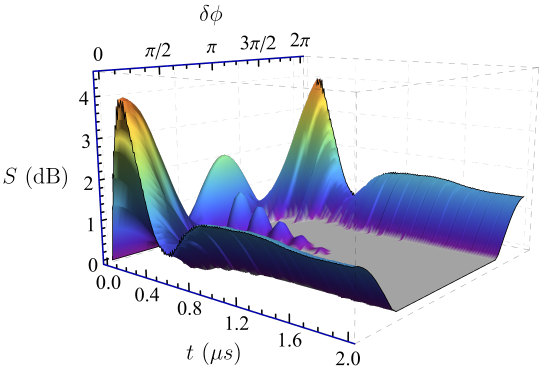}
	\caption{The effect of the relative phase between the two driving fields on the dynamics of the magnon squeezing. The gray area denotes $S \le 0$. The other parameters are the same as those in Fig.~\ref{fig:fig2}.} 
	\label{fig:fig5}
\end{figure}

\begin{figure}[t]
	\centering
	\includegraphics[width=0.9\linewidth]{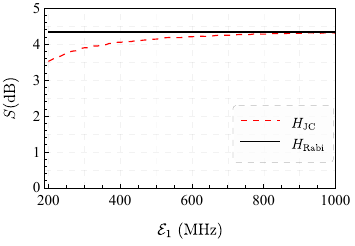}
	 {\caption{Maximum degree of squeezing versus Rabi frequency $\mathcal{E}_1/2\pi$. The red dashed curve is obtained by using the full driven JC Hamiltonian Eq.~\eqref{eq:jcprime}, while the black solid line corresponds to the effective $\mathcal{E}_1$-independent Rabi model Eq.~\eqref{eq:rabi}. }}
	\label{fig:fig6}
\end{figure}

\section{The effect of the relative phase between two driving fields  \label{appendixA}}

In writing the Hamiltonian Eq.~\eqref{eq:jcprime}, we assumed the initial phases of the two driving fields to be equal and zero, i.e., $\phi_1=\phi_2=0$.  However, in general the relative phase between the driving fields $\delta \phi = \phi_1-\phi_2$ may have an appreciable effect. To examine this possible effect on the magnon squeezing, we include the relative phase of the two driving fields into our model by adding a phase term $e^{-i \delta \phi}$ in the second last driving term in Eq.~\eqref{eq:jcprime}.  
 In Fig.~\ref{fig:fig5}, we plot the degree of squeezing $S$ (dB) as a function of time and the relative phase $\delta\phi \in [0,2\pi]$. It reveals that  the relative phase indeed can significantly affect the squeezing: the optimal situation corresponds to the in-phase driving $\delta\phi = 0$ ($2\pi$), i.e., the case we considered in the main text, while other phase differences yield a reduced degree of squeezing.

 {\section{Analysis of the validity of the RWA for the effective Rabi model \label{appendixB}}}

 {Here, we numerically assess the validity of the RWA used to derive the effective Rabi Hamiltonian Eq.~\eqref{eq:rabi} from the full driven JC model Eq.~\eqref{eq:jcprime}. Specifically, in Fig.~6 we numerically solve the master equation Eq.~\eqref{eq:me} for the JC Hamiltonian Eq.~\eqref{eq:jcprime} at different values of the Rabi frequency $\mathcal{E}_1$ and achieve the maximum degree of squeezing at an optimal time. We then compare the results with those obtained by solving the master equation Eq.~\eqref{eq:me2} for the effective Rabi Hamiltonian Eq.~\eqref{eq:rabi}.  Clearly, the approximation, i.e., the Rabi model, is more accurate as $\mathcal{E}_1$ increases, which better satisfies the condition of the RWA. Even for a relatively small value $\mathcal{E}_1/2\pi = 200$ MHz, the deviation of the degree of squeezing is less than 1 dB, i.e., $  4.35 - 3.55 = 0.8$ dB. 
In plotting Fig.~6, we only vary the Rabi frequency $\mathcal{E}_1$ and accordingly the drive frequency $\omega_2$, such that $\omega_1 - \omega_2 = 2\mathcal{E}_1$ is always kept. 
The other parameters are the same as those in Fig.~\ref{fig:fig2}. }


%

\end{document}